\documentclass[12pt]{article}
\usepackage[margin=1in]{geometry}
\usepackage{graphicx}
\usepackage{authblk}
\usepackage{amsmath}
\usepackage{cite}
\usepackage{hyperref}

\title{High-$\beta^*$ Optics Calculus at IP2 for Forward Physics in LHC Runs 3 and 4}

\author[1]{Sorina Popescu}

\affil[1]{IFIN-HH, Bucharest, Romania }
\affil[2]{Institut für Kernphysik, Münster, Germany}
\date{May 2025}

\begin{document}
\maketitle

\begin{abstract}
We present a study of high-$\beta^*$ optics configurations at Interaction Point 2 (IP2) of the Large Hadron Collider (LHC), developed to enable forward and diffractive physics measurements with the ALICE experiment during Runs 3 and 4. Using MAD-X, we designed a $\beta^* = 30$~m optics scheme that satisfies beam stability and aperture requirements, while offering improved sensitivity to small-angle scattering. The configuration follows the Achromatic Telescopic Squeeze (ATS) optics scheme—originally developed for IP1 and IP5—which provides enhanced control over phase advance and chromaticity. The resulting optics layout enables a forward physics program with continuous data-taking. We also outline possible extensions toward even higher $\beta^*$ values and discuss the implementation roadmap.
\end{abstract}

\section{Introduction and Physics Motivation}

The forward region at the LHC provides unique access to both Standard Model (SM) and Beyond the Standard Model (BSM) phenomena through processes characterized by small scattering angles and large rapidity gaps. These include elastic and diffractive scattering, photon-induced reactions, and central exclusive production, all of which are highly sensitive to the underlying dynamics of proton and heavy-ion collisions.

From the SM perspective, forward physics enables precise determinations of the total and elastic cross sections, studies of soft QCD and diffractive mechanisms, investigations of parton distribution functions (PDFs) at low-$x$, and tests of quantum electrodynamics (QED) via photon-photon fusion. These processes are crucial for validating theoretical models and for improving cosmic ray shower simulations through measurements of forward particle production.

Forward detection also enhances the LHC’s sensitivity to BSM scenarios. Searches for anomalous quartic gauge couplings (aQGCs), Axion-Like Particles (ALPs), and long-lived particles (LLPs) benefit significantly from the reduced background and high resolution made possible by high-$\beta^*$ optics. These setups increase the transverse separation of scattered protons and reduce beam divergence, making it feasible to cleanly identify rare exclusive final states.

In heavy-ion ($PbPb$) collisions, the large electromagnetic fields associated with ultra-relativistic nuclei can give rise to photon-induced interactions such as jet photoproduction, coherent and incoherent vector meson production, and light-by-light scattering. These ultra-peripheral collisions (UPCs) produce low-$p_T$ final states with minimal hadronic overlap, ideal for ZDC-based tagging. The processes $\gamma + Pb \rightarrow J/\psi + Pb$ or $\gamma\gamma \rightarrow \ell^+\ell^-$ span invariant masses from a few GeV to over 100 GeV and probe the photon content and nuclear structure with unprecedented precision ~\cite{fp420, cms_aqgc, forward_lhc}.

In $pp$ collisions, diffractive and exclusive processes such as elastic scattering, single diffraction, central exclusive production (CEP), and $\gamma\gamma$ fusion are sensitive to the proton’s internal structure and to new physics. Proton tagging and forward detector coverage are critical for reconstructing the full event topology and for suppressing pile-up, especially in low-luminosity running scenarios.
\begin{figure}[h!]
\centering
\begin{minipage}[b]{0.60\textwidth}
  \centering
  \includegraphics[width=\textwidth]{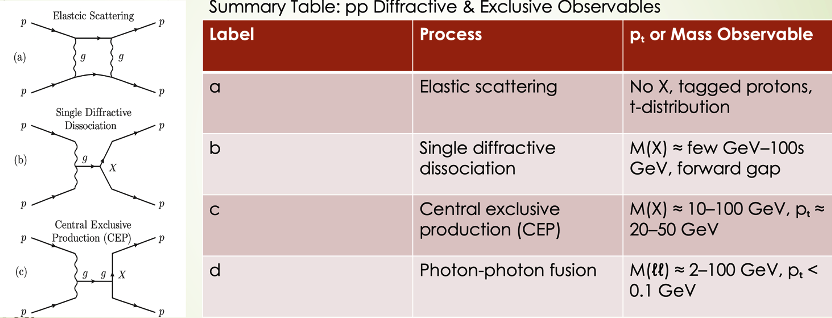}
  \caption{Forward physics observables in $pp$ collisions.}
  \label{fig:pp}  
\end{minipage}
\hfill
\begin{minipage}[b]{0.60\textwidth}
  \centering
   \includegraphics[width=\textwidth]{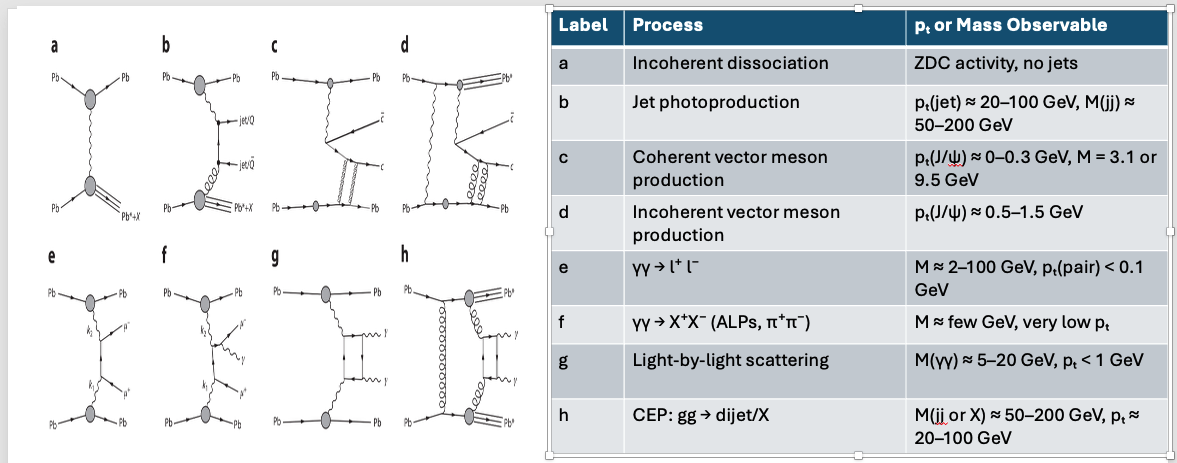}
   \caption{Heavy-ion observables accessible with forward detection~\cite{steinberg}.}
  \label{fig:PbPb}
\end{minipage}
\end{figure}

The deployment of a high-$\beta^*$ optics scheme at IP2, where ALICE is located, opens the door to a continuous forward physics program. Unlike previous special runs at IP1 and IP5, which were constrained to limited time windows and low statistics, the proposed configuration supports long-term data-taking with upgraded detectors. The key observables accessible in this program are summarized.

\section{Forward Detector Coverage and Physics Reach}

LHC experiments cover a wide range of pseudorapidities ($\eta$), but gaps exist. Rapidity coverage of LHC experiments is discussed in detail in ~\cite{forward_lhc}.

\begin{itemize}
  \item CMS/ATLAS: Central coverage to $|\eta| < 2.5$, forward calorimetry up to $|\eta| \approx 5$
 \item TOTEM/ALFA: Roman Pots covering $|\eta| > 6$
  \item LHCf: Zero-degree calorimeters for $|\eta| > 8$
\item ALICE: Forward muon spectrometers in $2 < \eta < 5$, optimized for heavy-ion and low pile-up pp physics
\end{itemize}

High-$\beta^*$ optics are essential to increase the separation of scattered protons and reduce divergence at the IP. This enhances the resolution for small-angle processes and suppresses pile-up, critical for low-mass central exclusive production.

\section{High-\texorpdfstring{$\beta^*$}{beta*} Optics Matching for IP2}

In order to fulfill the requirement of reduced luminosity and low divergence, a beam optics has been established with the optics code MAD-X (version 5.09) to provide matching solutions for IP2 optics with $\beta^* $ values up to $ 30$ m. This ``medium'' $\beta$ optics has been calculated for two scenarios, the standard LHC optics as well as the luminosity upgrade optics for the HL-LHC. Thus, the approach is compatible with the design version of the LHC as well as the so-called ATS optics ~\cite{ATS} as baseline for IP1 and IP5 in high-luminosity operation. 

\par
The matching of the new optics was performed in several steps, in order to guarantee a smooth transition from the LHC standard settings to the new low divergence optics. Successful magnet configurations leading to beta-functions up to $\beta*=30~m$ have been found for both beam 1 and beam 2. Within the given limits of the LHC magnet strengths (i.e. the maximum gradients allowed for safe beam operation) the goal of   $\beta*=30~m$ at full energy has been reached without problems. 
\begin{figure}[ht!]
\centering
\includegraphics[width=0.6\textwidth]{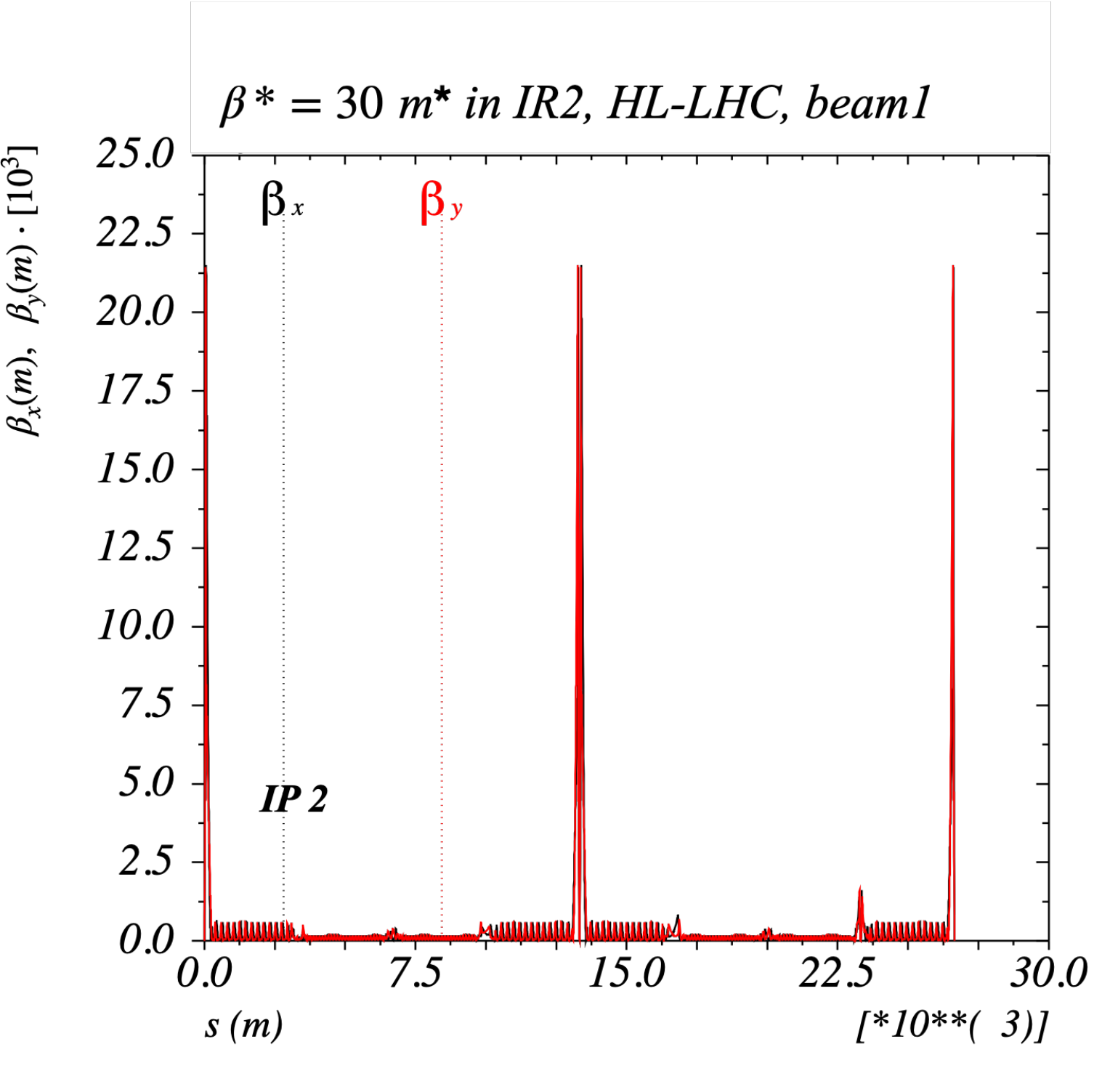}
\caption{Beam optics of LHC, beam1, for a low divergence  $\beta* = 30~m$ at IP2 embedded into the ATS-based HL-LHC optics in IP1 and IP5. On the left side of IP2 the optics has to fulfill the requirements of the ATS based beta wave while on the right side of IP2 the standard regular arc pattern is obtained. }
\label{beam1_all}
\end{figure}

\par
\begin{figure}[h!]
\centering
\includegraphics[width=0.6\textwidth]{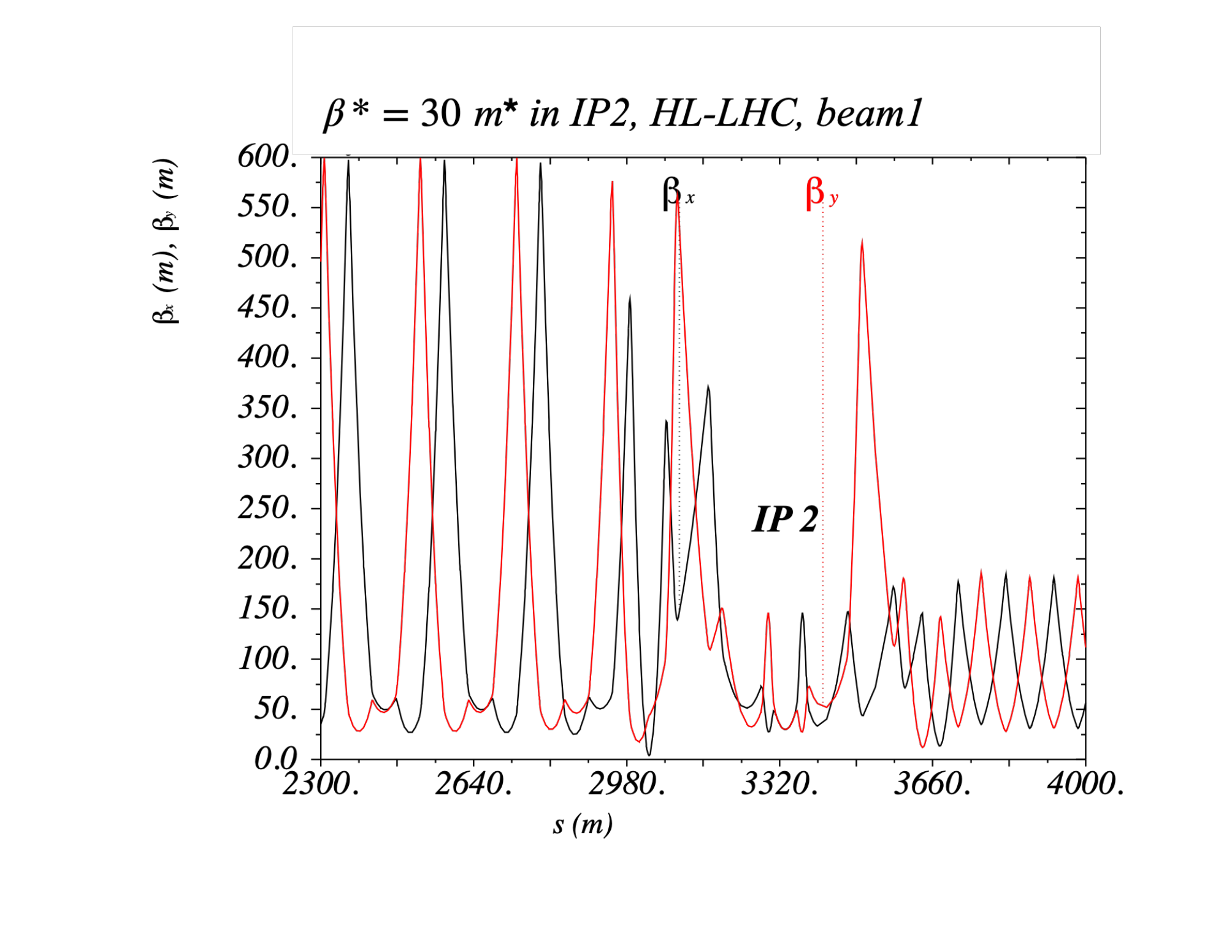}
\caption{Zoomed-in view of the new beam optics in IP 2. The beam size and so the aperture requirements in the straight section of IP2 are well within the boundaries set by the HL-LHC scheme.}
\label{beam1_IR2}
\end{figure}

 
The main goals of the matching were:
\begin{itemize}
  \item Achieve clean $\beta$-function profiles at IP2
  \item Ensure phase advance and chromatic stability
  \item create a new high beta optics IP 2 which is fully modular and can replace in straight forward manner the present IP 2 optics.
  \item Respect aperture limits and magnet constraints
\end{itemize}
Particular care was needed around the matching quadrupole MQM.6L2.B2, to avoid aperture limits due to eventually increased beam amplitudes. Final checks showed no showstoppers.
Fig.\,\ref{beam1_all} shows the successfully matched optics for LHC beam 1 with $\beta*= 30 m$ in IP2 and the high luminosity beam optics in IP1 and IP5. For better resolution, a zoomed-in view of this low-divergence optics is shown in Fig.\,\ref{beam1_IR2}. The matched connection to the ATS high-beta wave on the left side towards IP1 is clearly visible, as well as to the standard arc optics of the LHC on the right hand side towards IR3.

\begin{figure}[h!]
\centering
\includegraphics[width=0.6\textwidth]{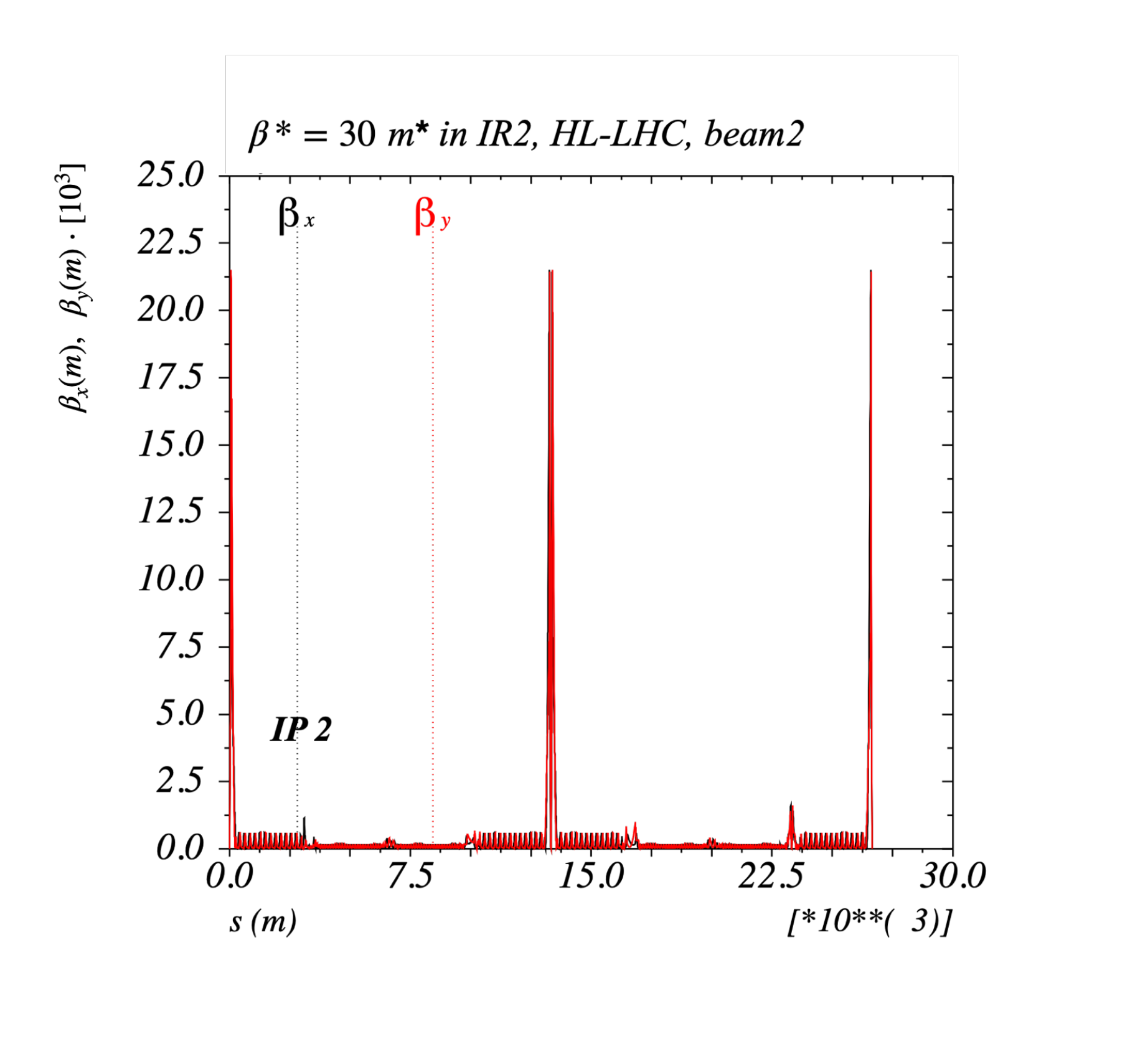}
\caption{Beam optics of LHC, beam2, for a low divergence beam of $\beta* = 30~m$ at IP2 embedded into the ATS-based HL-LHC optics in IR1 and IR5. }
\label{beam2_all}
\end{figure}

For completeness we add the optics matched for LHC beam 2, combining as before the low divergence settings in IP2, with $\beta*=30~m$, with the high luminosity optics in IP1 and IP5, Fig.\,\ref{beam2_all}
\par
Due to the complete anti-symmetric layout of the lattice of beam 1 and beam 2 of the LHC perfect matching between the beam parameters at the interaction point IP2 has been obtained. 
Further exploration suggests the optics can be pushed even further and values up to  $\beta^* = 35$ or even $40 m$ should be in reach, depending on further optimisation and LMC approval. Preliminary checks of quadrupole strengths indicate the technical feasibility, mainly defined by the quench safety margin of the quadrupole gradients.
\par 
The transition from the injection optics to the new low divergence collision optics has been set up through a series of about 12 intermediate beam optics. They guarantee a smooth transition from injection energy, where the magnet settings are mainly determined by the large beam emittance and the corresponding aperture restrictions, to the new collision case described above. The same calulus has been performed for the Run 3 optics and results are also matching.

\section{Summary and Implementation Path}

The $\beta^* = 30$ m optics solution at IP2 is validated and machine-compatible, enabling ALICE to explore a forward physics program with continuous data-taking. The next steps for implementation include:

\begin{itemize}
  \item ALICE endorsement and formal optics request to the LHC Program Coordinator
  \item Safety checks (vacuum, cryo, magnet strength) coordinated by appointed experts
  \item Inclusion in Machine Development (MD) schedule—tentatively 2025 or 2026
\end{itemize}

This setup provides a long-awaited opportunity to probe diffractive dynamics and exclusive production with sufficient luminosity and detector resolution.
While this paper focuses on the optics design and implementation feasibility, a full physics case including detector integration, background suppression, and expected signal sensitivity will be the subject of future dedicated studies.

\section*{Acknowledgements}
We thank the CERN, BE-ABB experts P. Hermes, B. Holzer and R. De Maria for many valuable discussions and their technical support during the optics matching studies.  \\
This work was supported in part by the ExtreMe Matter Institute (EMMI) through the EMMI Award, granted by GSI Helmholtz Centre for Heavy Ion Research. The author gratefully acknowledges the support provided during the research period hosted at the Heidelberg Physics Institute and insightful discussions with J. Stachel and R. Schicker, whose experience in heavy-ion physics greatly enriched the conceptual framework of this work.

\bibliographystyle{unsrt}
\bibliography{refs}

\end{document}